\documentclass[12pt]{article}
\usepackage{graphics,graphicx,graphpap,color}
\usepackage{setspace}

\newcommand{\GeV}{~\mbox{GeV}}

\newcommand\bmat{\left( \begin{array}{cc}}
\newcommand\emat{\end{array}\right)}

\def\msbar{\ifmmode{\overline{\rm MS}} \else{$\overline{\rm MS}$} \fi}
\def\drbar{\ifmmode{\overline{\rm DR}} \else{$\overline{\rm DR}$} \fi}

\def\ti              {\tilde}

\def\a               {\alpha}
\def\b               {\beta}
\def\d               {\delta}
\def\D               {\Delta}

\def\g               {\gamma}

\def\t               {\theta}
\def\s               {\sigma}

\def\x               {\chi}

\def\sf              {{\ti f}}

\def\sfL             {{\ti f_L^{}}}
\def\sfR             {{\ti f_R^{}}}

\def\chp             {\ti \x^+}

\newcommand{\msf}[1]   {m_{\ti f_{#1}}}
\newcommand{\mst}[1]   {m_{\ti t_{#1}}}
\newcommand{\msb}[1]   {m_{\ti b_{#1}}}
\newcommand{\mstau}[1] {m_{\ti\tau_{#1}}}

\def\tw              {\t_{\scriptscriptstyle W}}
\def\tb              {\tan\beta}

\def\onehfbi         {{\displaystyle\frac{1}{2}}}

\def\gev             {{\rm GeV}}
\def\tev             {{\rm TeV}}

\def\non             {\nonumber}

\renewcommand\d{\delta}

\begin{document}

\pagestyle{empty} \vspace*{-1cm}

\begin{flushright}
  HEPHY-PUB 784/04 \\
  hep-ph/0401092
\end{flushright}

\vspace*{2cm}

\begin{center}
{\Large\bf\boldmath
   Complete one-loop corrections to {\boldmath{ $e^+ e^-
   \rightarrow \tilde{f}_i\ {\bar{\!\!\tilde{f}}}_{\!j}$}}}
   \\[5mm]

\vspace{10mm}

$\mbox{K.~Kova\v{r}\'{\i}k}^{1,2}, \mbox{C.~Weber}^1,
\mbox{H.~Eberl}^1, \mbox{W.~Majerotto}^1$\\[5mm]

\vspace{6mm} $$\phantom{a}^1 \mbox{{\it Institut f\"ur
Hochenergiephysik der \"Osterreichischen Akademie der
Wissenschaften,}}$$\vspace{-0.9cm} $$\mbox{{\it A-1050 Vienna,
Austria}}$$ $$\phantom{a}^2 \mbox{{\it Department of Theoretical
Physics FMFI UK, Comenius University,}}$$\vspace{-0.65cm}
$$\mbox{{\it SK-84248 Bratislava, Slovakia}}$$
\end{center}

\vspace{20mm}

\begin{abstract}
We have calculated the complete one-loop corrections to the
sfermion pair production process $e^+ e^- \rightarrow \tilde{f}_i\
{\bar{\!\!\tilde{f}}}_{\!j}\;(f = t, b, \tau, \nu_\tau)$ in the
Minimal Supersymmetric Standard Model. Our results also include
the previously calculated SUSY-QCD corrections. We present the
details of the renormalization scheme used. It is found that the
weak corrections are of the same magnitude as the SUSY-QCD
corrections at higher energies ($\sqrt{s}\sim 1\tev$). At these
energies an important part of the weak corrections stems from the
box contribution. This is best seen in sneutrino production.
\end{abstract}

\vfill
\newpage
\pagestyle{plain} \setcounter{page}{2}

\section{Introduction}
\vspace{2mm} If supersymmetry (SUSY) is realized in Nature there
should be two scalar particles (sfermions) $\sf_L$, $\sf_R$
corresponding to the two chirality states of each fermion $f$. The
sfermions of the third generation play a special role as $\sf_L$
and $\sf_R$ may strongly mix (proportionally to the fermion mass),
forming the two mass eigenstates $\sf_1$ and $\sf_2$ (with $f= t,
b, \tau$) . As a consequence one eigenstate ($\sf_1$) can have  a
much lower mass than the other one. \newline Sfermion pair
production in $e^+e^-$ collisions, $e^+e^- \rightarrow
\tilde{f}_i\,\ {\bar{\!\!\tilde{f}}}_{\!j}, (i,j=1,2)$, has been
studied extensively phenomenologically \cite{exp}. The strong
interest in sfermion production is mainly due to the fact that it
gives access to one of the fundamental SUSY breaking parameters
$A_f$, the trilinear coupling parameter. It is clear that in
$e^+e^- \rightarrow \tilde{t}_i\,\ {\bar{\!\!\tilde{t}}}_{j}$ and
$\tilde{b}_i\,\ {\bar{\!\!\tilde{b}}}_{j}$  gluon radiation and
gluon exchange play an important role \cite{QCD1, QCD2}. The
SUSY-QCD corrections to these processes due to gluino and squark
exchange were calculated in \cite{SUSY-QCD-A, SUSY-QCD-H} and
found to become effective at $\sqrt{s} > 500 \GeV$. Yukawa
corrections \cite{Yukawa} were shown to be non negligible either.
Very recently, while we were already working on the calculation of
the full one-loop corrections to sfermion pair production within
the Minimal Supersymmetric Standard Model (MSSM), such a
calculation was presented in \cite{hollik}. In this context, it is
worthwhile to mention that the complete one-loop corrections to
selectron and smuon pair production from threshold to high
energies were calculated in \cite{freitas}.\newline For the
calculation of higher order corrections, renormalization of the
MSSM  with an appropriate fixing of the SUSY parameters is
necessary. Essentially, two methods were proposed in the on-shell
scheme, one in \cite{scheD1, scheD2} and the other one in
\cite{sche1, sche2} for a review see  \cite{sche3}. Of course,
both should lead to the same results for observables as masses,
cross-sections, widths , etc.. The schemes differ in the fixing of
the counterterms  of some of the SUSY parameters as $M_1, M_2,
\mu$, etc. Therefore the meaning of these parameters is different
at loop level. However, at one-loop, in sfermion pair production
this difference only matters in selectron pair production, $e^+e^-
\rightarrow \tilde{e}_i\ {\bar{\!\tilde{e}}}_{j}$ , and sneutrino
pair production, $e^+e^- \rightarrow \tilde{\nu}_e\
{\bar{\!\tilde{\nu}}}_{e}$ , due to the neutralino or chargino
exchange being already present at tree level. Here we fix the SUSY
parameters entering the sfermion mass matrices in the
corresponding sfermion sector so that one has not to take into
account any shifts in these parameters at one-loop level. In our
case, we also have a different fixing of the fine structure
constant $\alpha$ taking $\alpha (m_Z)$ as input, in contrast to
\cite{hollik} where the Thomson limit is used according to
\cite{Denner}.
\newline In this paper, we also present a full one-loop calculation
within the MSSM for $e^+e^- \rightarrow \tilde{f}_i\
{\bar{\!\!\tilde{f}}}_{\!j}, f = t, b, \tau,\nu_{\tau}$. We
compare our results with those obtained in \cite{hollik}. Due to
the complexity of such a calculation, an independent computation
seems appropriate. We have calculated all graphs analytically and
have written our own computer program for the numerical
evaluation. Checks have been performed using the computational
package \cite{loopFF, feyn}. In addition to the comparison with
\cite{hollik}, we studied different physical scenarios in our
numerical analysis. We also include a study of the tau-sneutrino
production which was not presented elsewhere.

\section{Tree level}\label{treelevel}
The sfermion mixing is described by the sfermion mass matrix in
the left-right basis $(\sfL, \sfR)$, and in the mass basis
$(\sf_1, \sf_2)$, $f = t,b$ or $\tau$ \cite{GunionHaber1,
GunionHaber2},
\begin{eqnarray}
  {\cal M}_{\sf}^{\,2} \,=\,
   \left(
     \begin{array}{cc}
       m_{\sf_L}^{\,2} & a_f\, m_f
       \\[2mm]
       a_f\,m_f & m_{\sf_R}^{\,2}
     \end{array}
   \right)
  = \left( R^\sf \right)^\dag
   \left(
     \begin{array}{cc}
       m_{\sf_1}^{\,2} & 0
       \\[2mm]
       0 & m_{\sf_2}^{\,2}
     \end{array}
   \right) R^\sf \,,
\end{eqnarray}
where $R^\sf_{i\a}$ is a 2 x 2 rotation matrix with rotation angle
$\theta_{\sf}$, which relates the mass eigenstates $\sf_i$, $i =
1, 2$, $(m_{\sf_1} < m_{\sf_2})$ to the weak eigenstates
$\sf_\a$, $\a = L, R$, by $\sf_i = R^\sf_{i\a} \sf_\a$ and
\begin{eqnarray}
  m_{\sf_L}^{\,2} &=& M_{\{\ti Q,\,\ti L \}}^2
       + (I^{3L}_f \!-\! e_f^{}\sin^2\!\tw)\cos2\b\,
       m_{\scriptscriptstyle Z}^{\,2}
       + m_{f}^2\,, \\[2mm]\label{MsD}
  m_{\sf_R}^{\,2} &=& M_{\{\ti U,\,\ti D,\,\ti E \}}^2
       + e_{f}\sin^2\!\tw \cos2\b\,m_{\scriptscriptstyle Z}^{\,2}
       + m_f^2\,, \\[2mm]
  a_f &=& A_f - \mu \,(\tan\b)^{-2 I^{3L}_f} \,.
\end{eqnarray}
$M_{\ti Q}$, $M_{\ti L}$, $M_{\ti U}$, $M_{\ti D}$ and $M_{\ti E}$
are soft SUSY breaking masses, $A_f$ is the trilinear scalar
coupling parameter, $\mu$ the higgsino mass parameter, $\tan\b =
\frac{v_2}{v_1}$ is the ratio of the vacuum expectation values of
the two neutral Higgs doublet states , $I^{3L}_f$ denotes the
third component of the weak isospin of the fermion $f$, $e_f$ the
electric charge in terms of the elementary charge $e$, and $\tw$
is the Weinberg angle.
\\
The mass eigenvalues and the mixing angle in terms of the primary
parameters are
\begin{eqnarray}
  \msf{1,2}^2
    &=& \frac{1}{2} \left(
    \msf{L}^2 + \msf{R}^2 \mp
    \sqrt{(\msf{L}^2 \!-\! \msf{R}^2)^2 + 4 a_f^2 m_f^2}\,\right)
\\
  \cos\t_{\sf}
    &=& \frac{-a_f\,m_f}
    {\sqrt{(\msf{L}^2 \!-\! \msf{1}^2)^2 + a_f^2 m_f^2}}
  \hspace{2cm} (0\leq \t_{\sf} < \pi) \,,
\end{eqnarray}
and the mass of the sneutrino $\nu_\tau$ is given by
$m_{\nu_\tau}^2 = M_{\ti L}^2 + \frac{1}{2}\,m_Z^2 \cos2\beta$.
\newline
The tree-level cross-section of $e^+e^- \rightarrow \tilde{f}_i \
{\bar{\!\!\tilde{f}}}_{\!j}$ is given by
\begin{eqnarray}
\s^{\rm tree}(e^+e^- \rightarrow \tilde{f}_i \
{\bar{\!\!\tilde{f}}}_{\!j}) &=& \frac{N_C}{3}\frac{\kappa^3 (s,
m^2_{\sf_i}, m^2_{\sf_j})}{16 \,\pi\, s^2}\left(T_{\g\g}+T_{\g
Z}+T_{ZZ}\right)\,,
\end{eqnarray}
where
\begin{eqnarray}
T_{\g\g}&=&\frac{e^4 e_f^2
(\d_{ij})^2}{s^2}\onehfbi(K_L^2+K_R^2)\,,
\\
T_{\g Z}&=& -\frac{g_Z^2 e^2 e_f a_{ij}^\sf\d_{ij}}{4s
(s-m_Z^2)}(C_L K_L +C_R K_R)\,,
\\
T_{ZZ}&=&\frac{g_Z^4(a_{ij}^\sf)^2}{32\,(s-m_Z^2)^2}(C_L^2+C_R^2)\,,
\end{eqnarray}
and $\kappa (x, y, z) = \sqrt{(x-y-z)^2 - 4 y z}$.\newline Here we
use $K_{L,R}$ and $C_{L,R}$ as the left- and right-handed
couplings of the electron to the  photon and Z boson,
respectively,
\begin{equation}
K_L = K_R = 1,\qquad C_L = -\frac{1}{2}+s_W^2, \qquad C_R = s_W^2.
\end{equation}
The matrix elements $a_{ij}^\sf$ come from the coupling of
$Z\sf_{i}\sf_{j}$,
\begin{equation}
a_{ij}^\sf = \left(\begin{array}{cc} 4(I^{3L}_f \cos^2\t_\sf-s_W^2
e_f) & -2I^{3L}_f \sin 2\t_\sf \\ -2I^{3L}_f \sin 2\t_\sf &
4(I^{3L}_f \cos^2\t_\sf-s_W^2 e_f)
\end{array}\right).
\end{equation}

\section{Radiative corrections}\label{radcor}
\begin{figure}[th]
\begin{picture}(160,215)(0,0)
     \put(0,-2){\mbox{\resizebox{15.5cm}{!}
     {\includegraphics{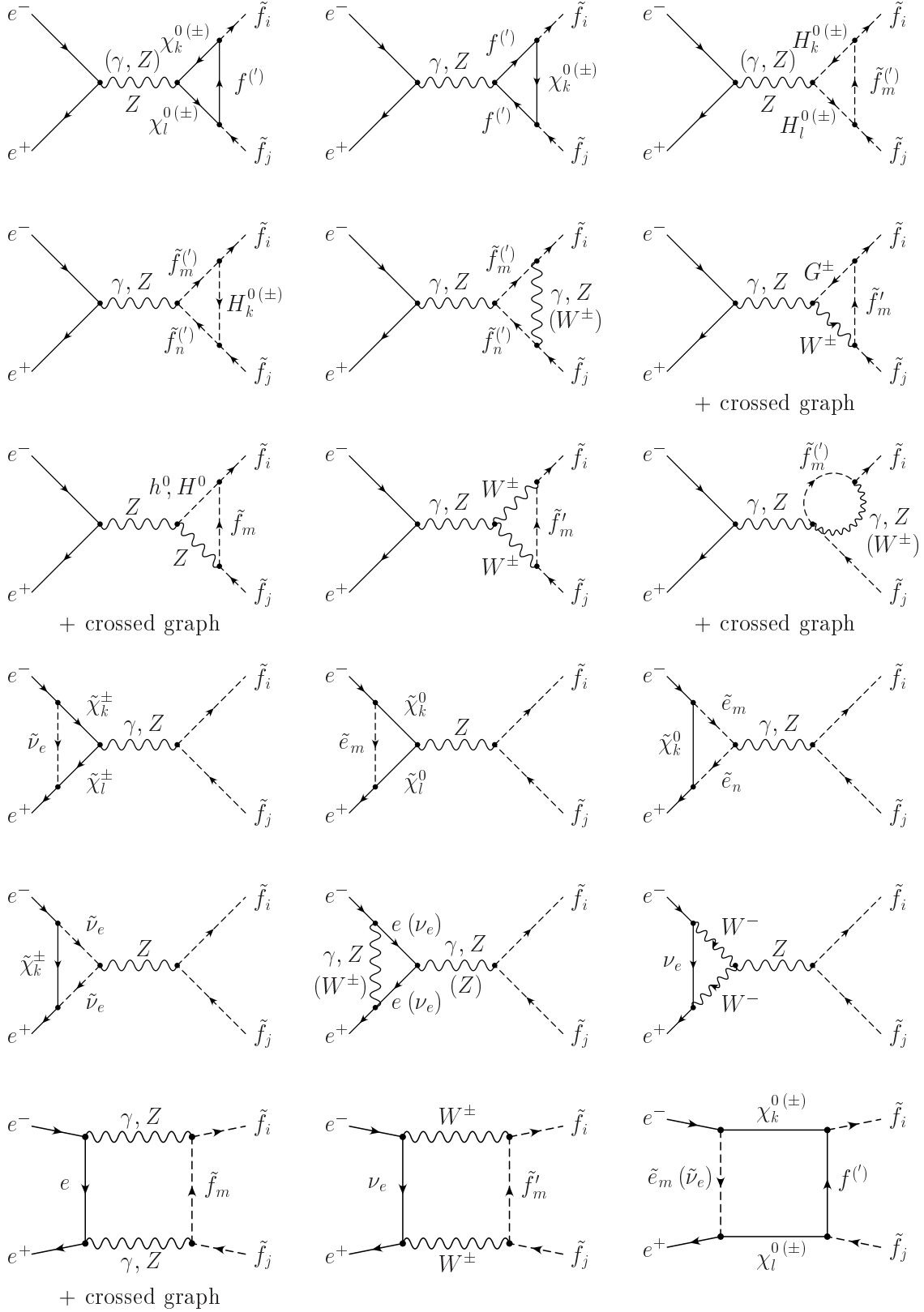}}}}
\end{picture}
\caption{Vertex and box diagrams relevant to the calculation of
the electroweak corrections to $e^+ e^- \rightarrow \tilde{f}_i \
{\bar{\!\!\tilde{f}}}_{\!j}$. \label{vertex-graphs}}
\end{figure}

\begin{figure}[th]
\begin{picture}(170,180)(0,0)
     \put(0,0){\mbox{\resizebox{15.5cm}{!}
     {\includegraphics{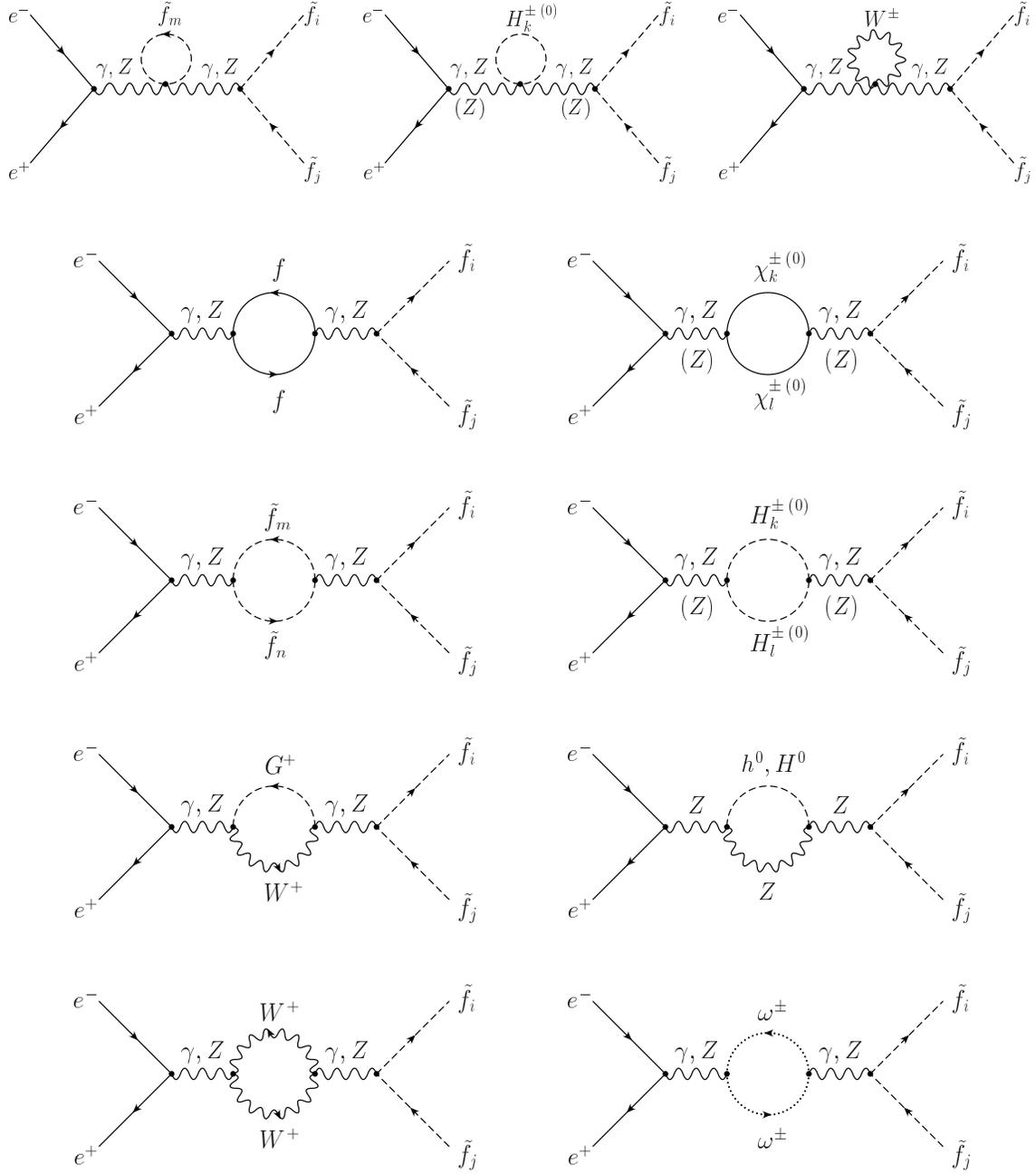}}}}
\end{picture}
\caption{Propagators relevant to the calculation of the
electroweak corrections to $e^+ e^- \rightarrow \tilde{f}_i \
{\bar{\!\!\tilde{f}}}_{\!j}$. \label{props}}
\end{figure}

The one-loop (renormalized) cross-section $\sigma^{\rm ren}$ is
expressed as
\begin{equation}
\s^{\rm ren}(e^+e^- \rightarrow \tilde{f}_i \
{\bar{\!\!\tilde{f}}}_{\!j})=\s^{\rm tree}+\D\s^{\rm
QCD}+\D\s^{\rm EW},
\end{equation}
where the symbol $\D$ denotes UV-finite quantities.\newline As
mentioned in the Introduction, the SUSY-QCD corrections
($\D\s^{\rm QCD}$) has already been calculated. In this paper, we
give the result for the complete one-loop electroweak corrections
($\D\s^{\rm EW}$) using the SUSY invariant dimensional reduction
($\overline{\rm DR}$) as the regularization scheme. The
calculation was performed in the 't Hooft-Feynman gauge, $\xi =
1$.\newline The electroweak corrections can be split further into
the following UV-finite parts as:
\begin{equation}
\D\s^{\rm EW} = \D\s^{\rm V\hspace{1pt}e}+\D\s^{\rm
V\hspace{1pt}\tilde{f}}+\D\s^{\rm prop}+\D\s^{\rm box},
\end{equation}
where $\D\s^{\rm V\hspace{1pt}e}$ and $\D\s^{\rm
V\hspace{1pt}\tilde{f}}$ stand for the renormalized electron and
sfermion vertex, $\D\s^{\rm prop}$ for renormalized propagators
 and $\D\s^{\rm box}$ for the box
contribution.\newline The renormalized electron vertex has the
form
\begin{eqnarray}
\D\s^{\rm V\hspace{1pt}e} &=& \frac{N_C}{3}\frac{\kappa^3 (s,
m^2_{\sf_i}, m^2_{\sf_j})}{16 \,\pi\, s^2}\left(\D
T^{V\hspace{1pt}e}_{\g\g}+\D T^{V\hspace{1pt}e}_{\g Z}+\D
T^{V\hspace{1pt}e}_{ZZ}\right),
\end{eqnarray}
where
\begin{eqnarray}
\D T^{V\hspace{1pt}e}_{\g\g}&=&\frac{e^4
e_f^2(\d_{ij})^2}{s^2}\,(\D e_L K_L+\D e_R K_R),\\ \D
T^{V\hspace{1pt}e}_{\g Z}&=& -\frac{g_Z^2 e^2 e_f
a_{ij}^\sf\d_{ij}}{4s (s-m_Z^2)}\,(\D e_L C_L +\D e_R C_R +\D a_L
K_L +\D a_R K_R ),\\ \D
T^{V\hspace{1pt}e}_{ZZ}&=&\frac{g_Z^4(a_{ij}^\sf)^2}{16\,(s-m_Z^2)^2}\,(\D
a_L C_L+ \D a_R C_R).
\end{eqnarray}
$\D e_{L,R}$ and $\D a_{L,R}$ consist of 3 parts,
\begin{eqnarray}
\D e_{L,R} &=& \d e_{L,R}^{(v)}+\d e_{L,R}^{(w)}+\d
e_{L,R}^{(c)}\,,\\ \D a_{L,R} &=& \d a_{L,R}^{(v)}+\d
a_{L,R}^{(w)}+\d a_{L,R}^{(c)}\,.
\end{eqnarray}
$\d e_{L,R}^{(v)}$, $\d a_{L,R}^{(v)}$ correspond to the vertex
corrections in Fig.~\ref{vertex-graphs}, $\d e_{L,R}^{(w)}$, $\d
a_{L,R}^{(w)}$ are the wave-function corrections (Fig.
\ref{props}) and $\d e_{L,R}^{(c)}$, $\d a_{L,R}^{(c)}$ correspond
to the counterterms.\newline The renormalized sfermion vertex has
a similar form,
\begin{eqnarray}
\D\s^{\rm V\hspace{1pt}\tilde{f}} &=& \frac{N_C}{3}\frac{\kappa^3
(s, m^2_{\sf_i}, m^2_{\sf_j})}{16 \,\pi\, s^2}\left(\D
T^{V\hspace{1pt}\tilde{f}}_{\g\g}+\D
T^{V\hspace{1pt}\tilde{f}}_{\g Z}+\D
T^{V\hspace{1pt}\tilde{f}}_{ZZ}\right)\,,
\end{eqnarray}
where
\begin{eqnarray}
\D T^{V\hspace{1pt}\tilde{f}}_{\g\g}&=&\frac{e^4 e_f (\D
e_f)_{ij}}{s^2}(K_L^2+K_R^2)\,,
\\
\D T^{V\hspace{1pt}\tilde{f}}_{\g Z}&=& -\frac{g_Z^2 e^2}{4s
(s-m_Z^2)}(K_L C_L + K_R C_R)\,((\D e_f)_{ij} a_{ij}^\sf +
\d_{ij}(\D a_f)_{ij})\,,
\\
\D T^{V\hspace{1pt}\tilde{f}}_{ZZ}&=&\frac{g_Z^4 a_{ij}^\sf (\D
a_f)_{ij}}{16\,(s-m_Z^2)^2}(C_L^2+ C_R^2)\,.
\end{eqnarray}
$(\D e_f)_{ij}$ and $(\D a_f)_{ij}$ can also be split into vertex
corrections (Fig.~\ref{vertex-graphs}), wave-function corrections
and counterterms,
\begin{eqnarray}
(\D e_f)_{ij} &=& (\d e_f)_{ij}^{(v)}+(\d e_f)_{ij}^{(w)}+(\d
e_f)_{ij}^{(c)}\,,\\ (\D a_f)_{ij} &=& (\d a_f)_{ij}^{(v)}+(\d
a_f)_{ij}^{(w)}+(\d a_f)_{ij}^{(c)}\,.
\end{eqnarray}
Explicit formulae for the remaining contributions $\D\s^{\rm
prop}$ and $\D\s^{\rm box}$ along with the formulae for
$e_{L,R}^{(v,w,c)},a_{L,R}^{(v,w,c)}(\d e_f)_{ij}^{(v,w,c)}, (\d
a_f)_{ij}^{(v,w,c)}$ will be given elsewhere.
\subsection{Fixing of the parameters}
The parameters occurring at the tree level, i.e. $m_W, m_Z,
\sin\tw$ are fixed as in \cite{Sirlin} and $\theta_{\!\sf}$ is
fixed according to \cite{SUSY-QCD-H, guasch}. The difference to
\cite{hollik} is our fixing of the electric charge since we use as
input parameter for $\a$ the $\overline{\rm MS}$ value at the
$Z$-pole, $\a \equiv \a(m_{\scriptscriptstyle Z})|_{\overline{\rm
MS}} = e^2/(4\pi)$. The counterterm then is \cite{chrislet,
0111303}
\begin{eqnarray}\non
  \frac{\d e}{e} &=& \frac{1}{(4\pi)^2}\,\frac{e^2}{6} \Bigg[
  \,4 \sum_f N_C^f\, e_f^2 \bigg(\D + \log\frac{Q^2}{x_f^2} \bigg)
  + \sum_{\sf} \sum_{m=1}^2 N_C^f\, e_f^2
  \bigg( \D + \log\frac{Q^2}{m_{\sf_m}^2} \bigg)
  \\
  && \hspace{18mm}
  +\,4 \sum_{k=1}^2 \bigg( \D + \log\frac{Q^2}{m_{\chp_k}^2} \bigg)
  + \sum_{k=1}^2 \bigg( \D + \log\frac{Q^2}{m_{H_k^+}^2} \bigg)
  - 22 \bigg( \D + \log\frac{Q^2}{m_{\scriptscriptstyle W}^2} \bigg)
  \Bigg] \,.\non
\end{eqnarray}
with $x_f = m_{\scriptscriptstyle Z} \ \forall\ m_f <
m_{\scriptscriptstyle Z}$ and $x_t = m_t$.  $N_C^f$ is the colour
factor, $N_C^f = 1, 3$ for (s)leptons and (s)quarks, respectively.
$\D$ denotes the UV divergence factor, \mbox{$\D = 2/\epsilon - \g
+ \log 4\pi$}.
\subsection{Real corrections}
\begin{figure}[th]
\begin{picture}(170,25)(0,0)
     \put(0,0){\mbox{\resizebox{16cm}{!}
     {\includegraphics{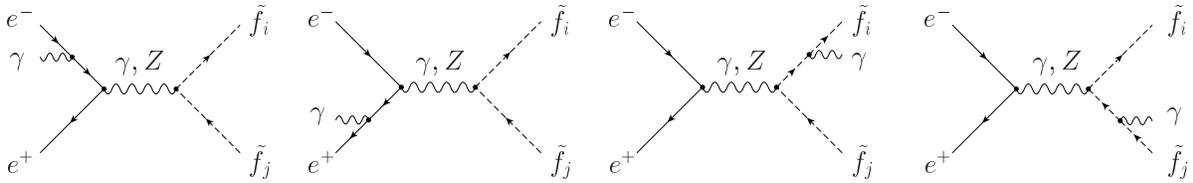}}}}
\end{picture}
\caption{Bremsstrahlung diagrams relevant to the calculation of
the real corrections to $e^+ e^- \rightarrow \tilde{f}_i \
{\bar{\!\!\tilde{f}}}_{\!j}$. \label{rads}}
\end{figure}

The cross-section $\s(e^+e^- \rightarrow \tilde{f}_i \
{\bar{\!\!\tilde{f}}}_{\!j})$ is IR-divergent owing to the photon
mass being zero. This is remedied by introducing a small mass
$\lambda$ and including also the Bremsstrahlung i.e. $\s(e^+e^-
\rightarrow \tilde{f}_i \ {\bar{\!\!\tilde{f}}}_{\!j}\g)$
(Fig.~\ref{rads}).
\clearpage

\clearpage \noindent

\noindent Summing those two contributions yields an IR-finite
result for the physical value $\lambda=0$,
\begin{equation}\label{radiat}
\s^{\rm corr}(e^+e^- \rightarrow \tilde{f}_i \
{\bar{\!\!\tilde{f}}}_{\!j})= \s^{\rm ren}(e^+e^- \rightarrow
\tilde{f}_i \ {\bar{\!\!\tilde{f}}}_{\!j})+\s(e^+e^- \rightarrow
\tilde{f}_i \ {\bar{\!\!\tilde{f}}}_{\!j}\g)\,.
\end{equation}
In our calculation, we used a soft-photon approximation
\cite{Denner} which reproduces the divergence pattern correctly
but introduces a cut $\D E$ on the energy of the radiated
photon.\newline The explicit formulae will be given elsewhere.

\section{Outline of the calculation}
The results presented in this paper come from a full analytic
calculation where we have neglected the electron mass ($m_e =0$)
except for the QED corrections.\newline In the case of this
particular process, we can separate off the QED corrections on the
basis of Feynman diagrams in a gauge invariant way. The QED
corrections consist of all the diagrams that have an additional
photon added to the tree level and therefore also include the
whole real corrections. The weak corrections are then UV and IR
finite and $\D E$ independent. All numerical results show only the
weak corrections as the QED part is very sensitive to $\D E$.
\newline
The numerical calculation was performed using the packages
LoopTools and FF \cite{loopFF}. The results were checked against
the results for the Yukawa approximation presented in
\cite{Yukawa} where our results match except for a minus sign
already pointed out in \cite{hollik}.\newline Furthermore, we did
our own independent calculation based on FeynArts and FormCalc
packages \cite{feyn} checking all individual Feynman diagrams at
the amplitude level. In addition, we used the packages to check
the complete results using the same renormalization scheme as in
the analytical calculation.\newline We also compared our results
with \cite{hollik} where we were able to reproduce all the results
apart from minor differences due to the different renormalization
of the fine structure constant.

\section{Numerical analysis}
\begin{figure}[bh!]
\begin{picture}(160,163)(0,0)
     \put(0,0){\mbox{\resizebox{15.5cm}{!}
     {\includegraphics{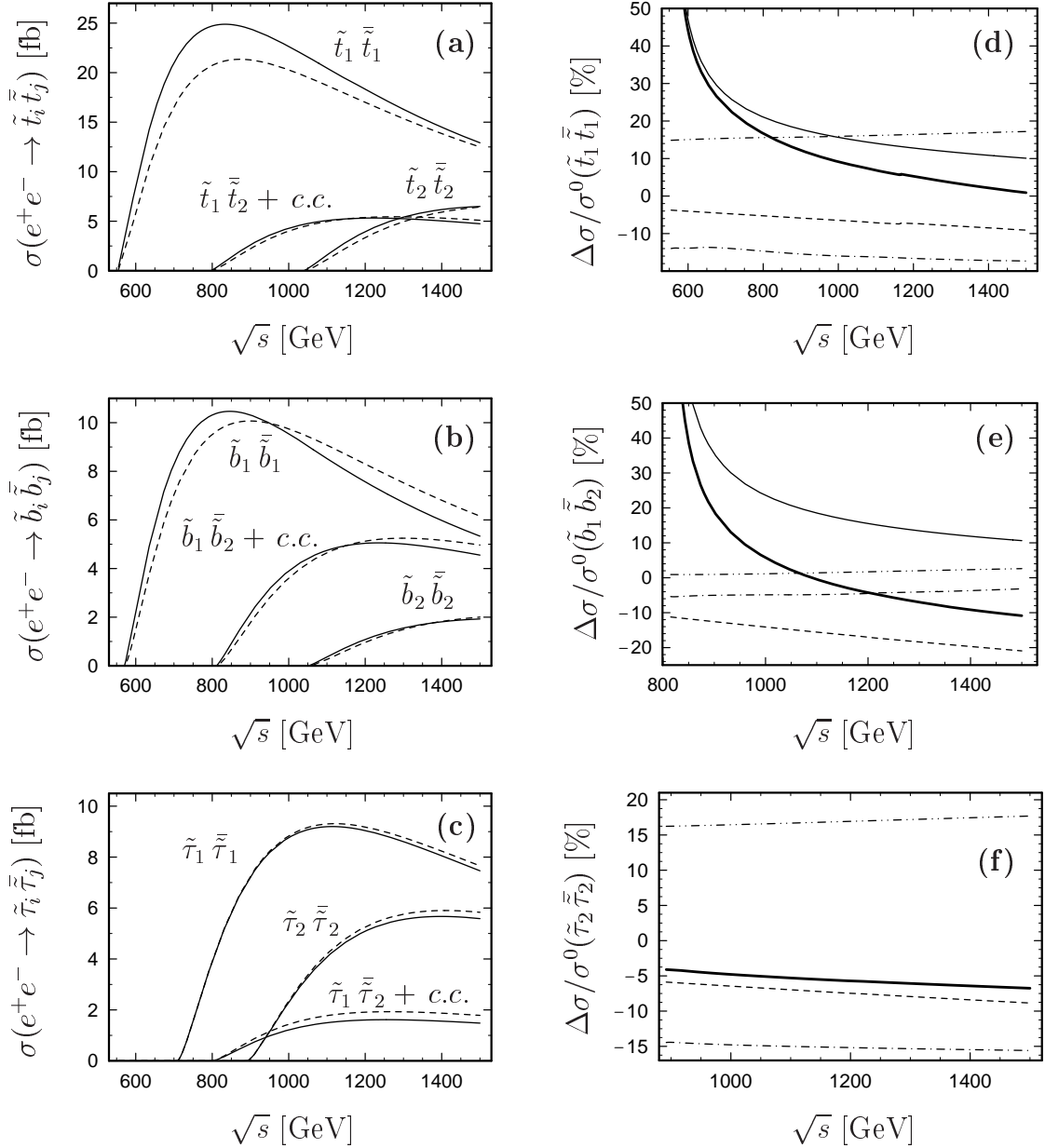}}}}
\end{picture}
\caption{Scenario 1: {\bf (a,b,c)} Total (solid) and tree-level
(dashed) cross-sections as a function of $\sqrt{s}$~; {\bf
(d,e,f)} Relative corrections to one of the channels from plots
{\bf (a,b,c)} split into convergent subclasses. (thick solid--
$\D\s^{\rm total}$, solid-- $\D\s^{\rm QCD}$, dashed-- $\D\s^{\rm
box}$, dash-dotted-- $\D\s^{\rm vertex}$, dash-dot-dotted--
$\D\s^{\rm prop}$)\label{sc1}}
\end{figure}

\begin{figure}[bh!]
\begin{picture}(160,163)(0,0)
     \put(0,0){\mbox{\resizebox{15.5cm}{!}
     {\includegraphics{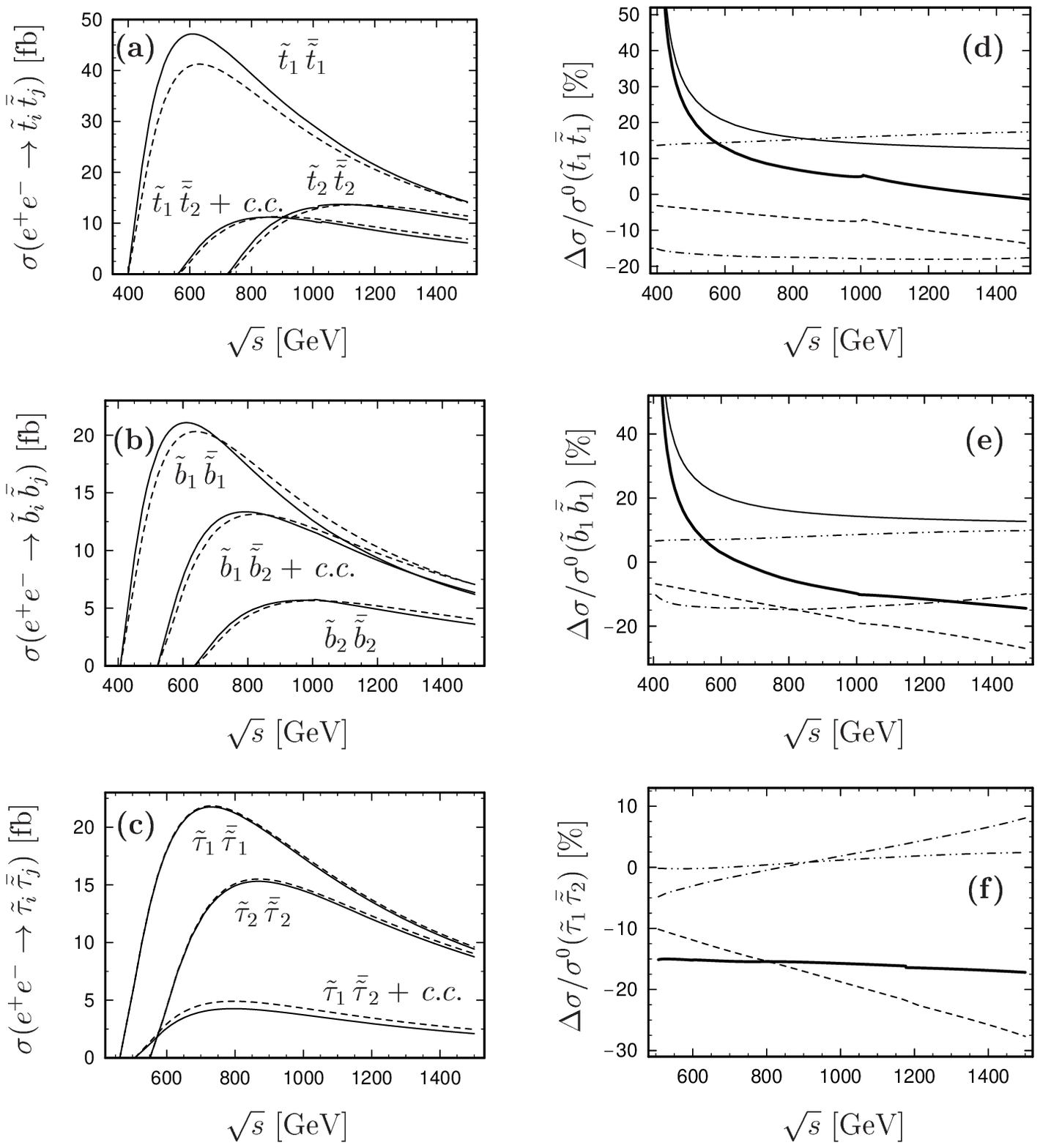}}}}
\end{picture}
\caption{Scenario 2: {\bf (a,b,c)} Total (solid) and tree-level
(dashed) cross-sections as a function of $\sqrt{s}$~; {\bf
(d,e,f)} Relative corrections to one of the channels from plots
{\bf (a,b,c)} split into convergent subclasses. (thick solid--
$\D\s^{\rm total}$, solid-- $\D\s^{\rm QCD}$, dashed-- $\D\s^{\rm
box}$, dash-dotted-- $\D\s^{\rm vertex}$, dash-dot-dotted--
$\D\s^{\rm prop}$)\label{sc2}}
\end{figure}

\begin{figure}[bh!]
\begin{picture}(160,163)(0,0)
     \put(0,0){\mbox{\resizebox{15.5cm}{!}
     {\includegraphics{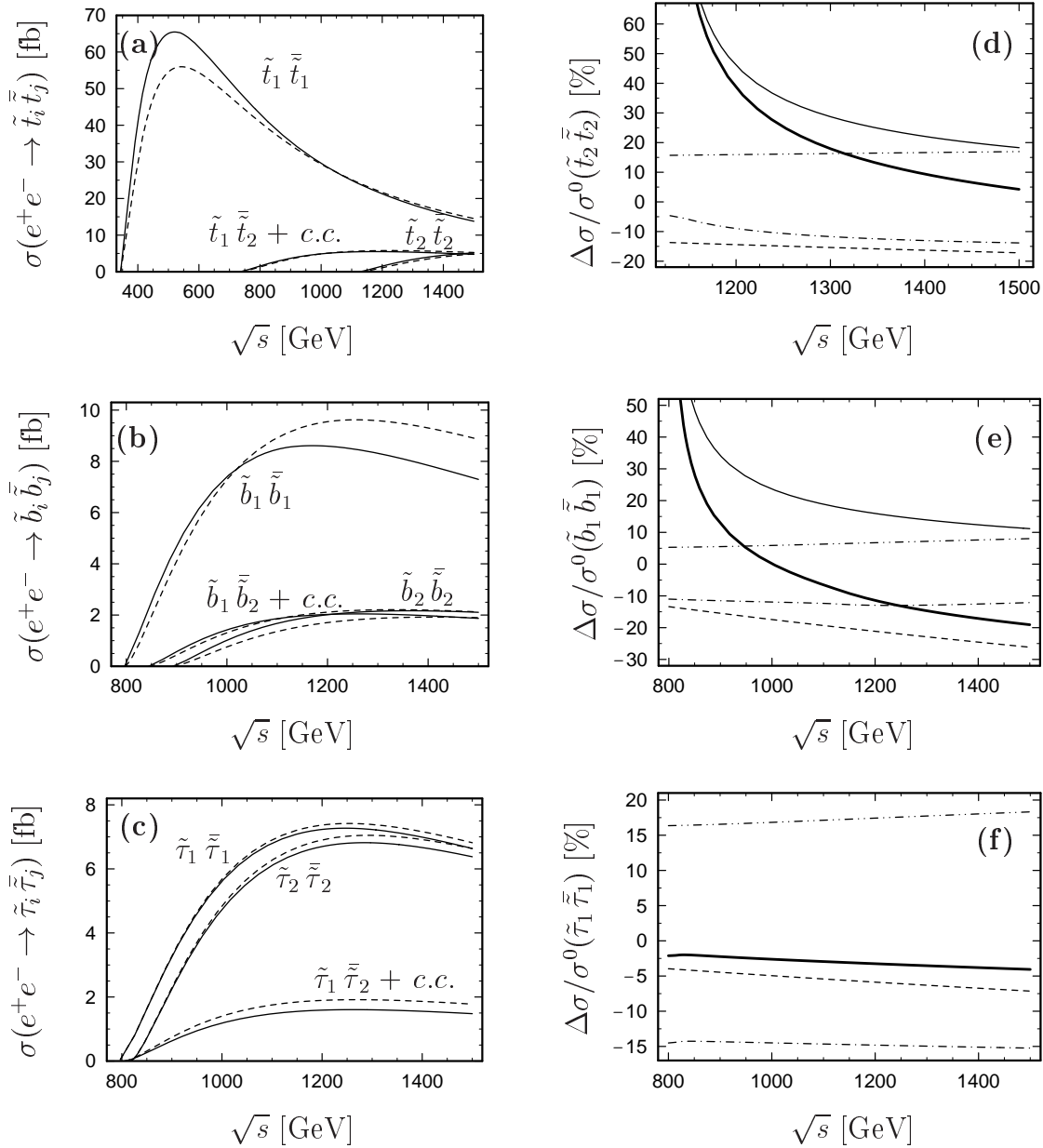}}}}
\end{picture}
\caption{Scenario 3: {\bf (a,b,c)} Total (solid) and tree-level
(dashed) cross-sections as a function of $\sqrt{s}$~; {\bf
(d,e,f)} Relative corrections to one of the channels from plots
{\bf (a,b,c)} split into convergent subclasses. (thick solid--
$\D\s^{\rm total}$, solid-- $\D\s^{\rm QCD}$, dashed-- $\D\s^{\rm
box}$, dash-dotted-- $\D\s^{\rm vertex}$, dash-dot-dotted--
$\D\s^{\rm prop}$)\label{sc3}}
\end{figure}

\begin{figure}[bh!]
\begin{picture}(160,170)(0,0)
     \put(0,0){\mbox{\resizebox{15.5cm}{!}
     {\includegraphics{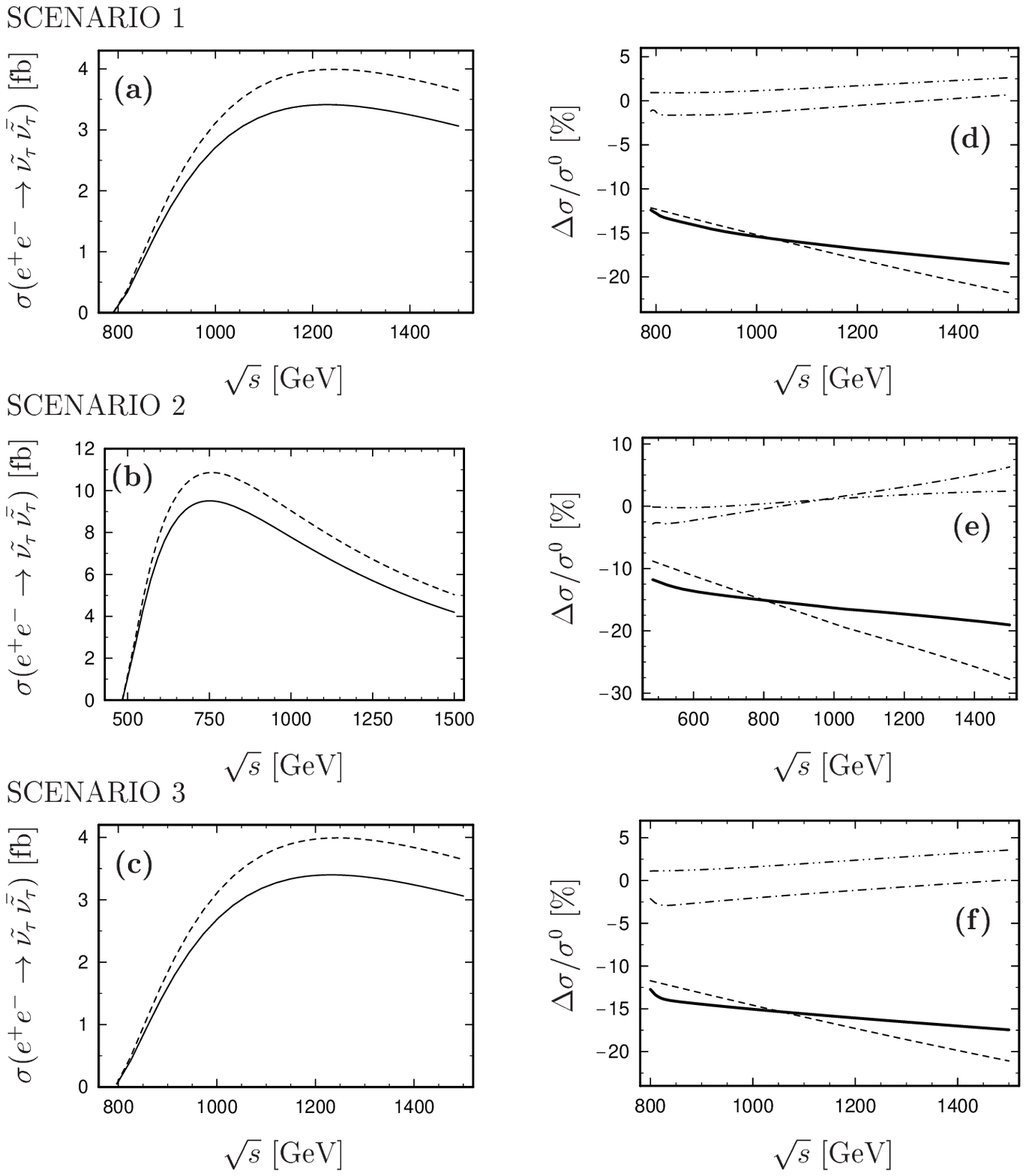}}}}
\end{picture}
\caption{Sneutrino plots: {\bf (a,b,c)} Total (solid) and
tree-level (dashed) cross-sections as a function of $\sqrt{s}$~;
{\bf (d,e,f)} Relative corrections split into convergent
subclasses.\hspace{3cm} (thick solid-- $\D\s^{\rm total}$,
dashed-- $\D\s^{\rm box}$, dash-dotted-- $\D\s^{\rm vertex}$,
dash-dot-dotted-- $\D\s^{\rm prop}$)\label{sne}}
\end{figure}

In the following numerical examples, we assume $M_{{\ti Q}} \equiv
M_{{\ti Q}_{3}} = \frac{10}{9} M_{{\ti U}_{3}} = \frac{10}{11}
M_{{\ti D}_{3}} = M_{{\ti L}_{3}} = M_{{\ti E}_{3}} = M_{{\ti
Q}_{1,2}} = M_{{\ti U}_{1,2}} = M_{{\ti D}_{1,2}} = M_{{\ti
L}_{1,2}} = M_{{\ti E}_{1,2}}$ for the first, second and third
generation soft SUSY breaking masses, and all trilinear couplings
are set to a common value $A$. For the standard model parameters
we take $m_{\scriptscriptstyle Z} = 91.1875$~GeV,
$m_{\scriptscriptstyle W} = 80.45$~GeV, $\sin^2 \tw = 1 -
m_{\scriptscriptstyle W}^2/m_{\scriptscriptstyle Z}^2$, $\a =
1/127.934$, $m_t = 174.3$~GeV, and $m_b = 4.7$~GeV. $M'$ is fixed
by the gaugino unification relation $M' =
{\displaystyle{\frac{5}{3}}} \, \tan^2\tw M$, and the gluino mass
is related to $M$ by $m_{\tilde g} = (\a_s(m_{\tilde
g})/\a)\sin^2\tw M$.\newline Below we show plots for three
different scenarios. On the left, there are the total and
tree-level cross-sections for all channels $e^+ e^- \rightarrow
\tilde{f}_i\ {\bar{\!\!\tilde{f}}}_{\!j},\, f=t,b,\tau,\, i=1,2$,
and on the right we picked out one of the channels and show a
separation into the convergent subclasses described in the text
above.\newline Care has to be taken when comparing plots for
different sfermions in the same scenario because we always fix the
SUSY parameters ($M_{{\ti Q}}, \mu$ etc.) in the sector of the
produced sfermions.
\newline One can see that the total corrections in squark
production are dominated by SUSY-QCD where the biggest
contribution comes from the gluon. The gluino part is small
compared to the gluon one as already shown in \cite{SUSY-QCD-H}.
At higher energy ($1~\tev-1.5~\tev$), other corrections can grow
to a size comparable to the SUSY-QCD contribution. The leading
weak contribution at high energy comes from the box diagrams. As
can be seen in Fig. \ref{sc1}-\ref{sc3}, the box diagrams give a
negative contribution rising with energy. This can be checked
against the Sudakov approximation at high energies
\cite{sudakov1,sudakov2}, where the box contributions give the
leading correction. The sum of the other two corrections, the
vertex and the propagator corrections, is small compared to the
box contribution.\newline The weak corrections computed here have
a significant effect in the high energy region, and in the case of
squarks act against the SUSY-QCD corrections. For stau production
they can go up to -10\% in certain cases.\newline In addition, we
also show the tau-sneutrino production. For detection, it is
necessary that not only the decay channel to $\chi^0_1$ is open.
Our three scenarios allow decays into charginos. Sneutrino
production is the only case where not only the SUSY-QCD
corrections are not present but also the Yukawa corrections are
small. This is due to the fact that the diagrams including a
neutralino or a neutral Higgs boson in the loop are missing (in
Yukawa approximation). Therefore, the sneutrino production shows a
particular dominance of the box corrections. In Fig. \ref{sne} one
sees that the tree level is almost identical in two of our
scenarios due to the small difference in the sneutrino mass. The
total cross-section in the two scenarios is also very similar as
the different vertex corrections and the propagator corrections
are together below 5\% and the boxes give the leading
contribution.
\section{Conclusion}
We have calculated the complete one-loop corrections to stop,
sbottom, stau and tau-sneutrino production. The calculation was
performed in an analytical way with an independent check using the
FeynArts and FormCalc computer packages \cite{feyn}. Our way of
fixing the fine structure constant $\alpha$ gives a higher
tree-level cross-section and therefore smaller radiative
corrections compared to \cite{hollik}. The corrections are
typically of 5-10\% and thus not negligible at a future linear
collider.
\newpage
\noindent {\bf SCENARIO 1- gaugino:}\newline\newline The
parameters are set to\newline $\{M,\mu,A,\tb,m_A,M_{{\ti
Q}}\}=\{200~\gev, 1000~\gev, -500~\gev, 20, 300~\gev,
400~\gev\}$\newline The masses of the sfermions in this scenario
are\newline $\mst{1,2}=\{276,520\}~\gev\quad\;
\msb{1,2}=\{285,526\}~\gev\quad\; \mstau{1,2}=\{354,446\}~\gev$
\\
\newpage
\noindent {\bf SCENARIO 2- higgsino:}\newline\newline The
parameters are set to\newline $\{M,\mu,A,\tb,m_A,M_{{\ti
Q}}\}=\{800~\gev, -200~\gev, 250~\gev, 30, 300~\gev,
250~\gev\}$\newline The masses of the sfermions in this scenario
are\newline $\mst{1,2}=\{200,360\}~\gev\quad\;
\msb{1,2}=\{203,318\}~\gev\quad\; \mstau{1,2}=\{231,275\}~\gev$
\\
\newpage
\noindent {\bf SCENARIO 3- mixed:}\newline\newline The parameters
are set to\newline $\{M,\mu,A,\tb,m_A,M_{{\ti Q}}\}=\{200~\gev,
200~\gev, -800~\gev, 10, 300~\gev, 400~\gev\}$\newline The masses
of the sfermions in this scenario are\newline
$\mst{1,2}=\{172,563\}~\gev\quad\;
\msb{1,2}=\{398,446\}~\gev\quad\; \mstau{1,2}=\{396,409\}~\gev$
\\
\newpage
\noindent {\bf SNEUTRINO PLOTS:}\newline\newline Here we show the
sneutrino production in all three scenarios. The masses of the
sneutrino in the scenarios are\newline
$m_{\ti\nu_\tau}=394.795~\gev\quad\;
m_{\ti\nu_\tau}=242~\gev\quad\; m_{\ti\nu_\tau}=394.873~\gev$
\\

\newpage
\noindent {\bf Acknowledgements}\\ \noindent The authors thank A.
Arhrib for communication. The authors acknowledge support from EU
under the HPRN-CT-2000-00149 network programme and the ``Fonds zur
F\"orderung der wissenschaftlichen Forschung'' of Austria, project
No. P13139-PHY.

\end{document}